\documentclass[sigconf]{acmart}

\AtBeginDocument{%
  }

\setcopyright{acmlicensed}
\copyrightyear{2018}
\acmYear{2018}
\acmDOI{XXXXXXX.XXXXXXX}

\acmConference[CIKM'24]{33rd ACM International Conference on
Information and Knowledge Management}{October 21--25, 2024}{Boise, Idaho, USA}
\acmISBN{978-1-4503-XXXX-X/18/06}

\usepackage{CJK}
\usepackage{multirow}
\usepackage{amsmath}
\usepackage{subcaption}
\usepackage{xcolor}
	
\usepackage{color, colortbl}
\definecolor{arsenic}{rgb}{0.23, 0.27, 0.29}
\definecolor{gray2}{gray}{0.9}

\begin{document}

\title{Towards Scalability and Extensibility of Query Reformulation Modeling in E-commerce Search}

\author{Ziqi Zhang}
\email{zzq@amazon.com}
\affiliation{%
  \institution{Amazon}
  \city{Beijing}
  \country{China}
}

\author{Yupin Huang}
\email{huayupin@amazon.com}
\affiliation{%
  \institution{Amazon}
    \city{Palo Alto}
  \country{USA}
}

\author{Quan Deng}
\email{quandeng@amazon.com}
\affiliation{%
  \institution{Amazon}
      \city{Seattle}
  \country{USA}
}

\author{Jinghui Xiao}
\email{xiaojhui@amazon.com}
\affiliation{%
  \institution{Amazon}
    \city{Beijing}
  \country{China}
}

\author{Vivek Mittal}
\email{vivekmit@amazon.com}
\affiliation{%
  \institution{Amazon}
      \city{Palo Alto}
  \country{USA}
}

\author{Jingyuan Deng}
\email{jingyua@amazon.com}
\affiliation{%
  \institution{Amazon}
    \city{Beijing}
  \country{China}
}

\renewcommand{\shortauthors}{Zhang, et al.}

\begin{abstract}
  Customer behavioral data significantly impacts e-commerce search systems. However, in the case of less common queries, the associated behavioral data tends to be sparse and noisy, offering inadequate support to the search mechanism. To address this challenge, the concept of query reformulation has been introduced. It suggests that less common queries could utilize the behavior patterns of their popular counterparts with similar meanings. In Amazon product search, query reformulation has displayed its effectiveness in improving search relevance and bolstering overall revenue. Nonetheless, adapting this method for smaller or emerging businesses operating in regions with lower traffic and complex multilingual settings poses the challenge in terms of scalability and extensibility. This study focuses on overcoming this challenge by constructing a query reformulation solution capable of functioning effectively, even when faced with limited training data, in terms of quality and scale, along with relatively complex linguistic characteristics. In this paper we provide an overview of the solution implemented within Amazon product search infrastructure, which encompasses a range of elements, including refining the data mining process, redefining model training objectives, and reshaping training strategies. The effectiveness of the proposed solution is validated through online A/B testing on search ranking and Ads matching. Notably, employing the proposed solution in search ranking resulted in 0.14\% and 0.29\% increase in overall revenue in Japanese and Hindi cases, respectively, and a 0.08\% incremental gain in the English case compared to the legacy implementation; while in search Ads matching led to a 0.36\% increase in Ads revenue in the Japanese case. The proposed solution has been deployed in Amazon product search.
\end{abstract}



\maketitle

\section{Introduction}
Search ranking and matching in E-commerce heavily rely on historical customer behavior like clicks and purchases. These actions genuinely reflect customer preferences for products associated with specific queries, offering valuable insights for refining the search system \cite{1}. The refinements based on customer behavior further enables the formation of a beneficial feedback loop that intertwines increased customer engagement with elevated search experience. However, challenges arise in activating this loop, particularly for queries with less popular or newly emerging shopping intents, which are categorized as "tail queries" \cite{goel2010anatomy}.

An intuitive strategy to overcome this challenge involves enabling tail queries to benefit from the behavioral data of more popular queries that convey similar semantic meanings \cite{2,3,4}. This approach is implemented through query reformulation (QR), which creates linkages between queries grounded in their semantic similarity. Despite substantial efforts to apply sophisticated techniques in optimizing QR solutions, there remains a gap in achieving the robustness required for industrial QR practices.

In the realm of E-commerce, we consider the robustness of a QR solution from two key angles: scalability and extensibility. Scalability pertains to the size and quality of data required to support the effective model training. It is widely recognized that having a sufficient amount of high-quality training data can significantly ease the intricacies of the training process and reduce system complexity. However, for smaller markets or niche business fields, historical customer engagement often tend to be both sparse and noisy, making it challenging to construct a high-quality training dataset. Extensibility decides whether the QR approach can be applied across a broad spectrum of linguistic scenarios, including those with more intricate text patterns.\footnote{take Japanese, for instance; unlike Indo-European languages, it comprises a mix of ideographic and phonetic characters, resulting in various script types, such as Kanji, Hiragana, Katakana, and Romaji, that may convey the same semantic meaning but are used in vastly different ways in diverse contexts} As such, our central focus in this paper is understanding how to build a QR solution capable of functioning effectively, even when faced with limited training data, in terms of quality and scale, along with relatively complex linguistic characteristics.


The primary challenge in achieving robustness of QR stems from its reliance on historical customer engagement - the development of the QR framework is susceptible to noise in training/evaluation data caused by uneven query exposure to customers \cite{szpektor2011improving}, evident in the form of false positive/negative query-query pairs. This noise are particularly prevalent with cold-start queries and those with limited association to related products due to system defects, which are exactly part of QR's target. In light of this challenge, we have refined core concepts and components of the legacy reformulation modeling \cite{2}. Our objective is to extend the benefit of the QR framework to smaller, non-English markets in addition to expand the benefit in major English markets. This paper presents a comprehensive overview of these improvements. Specifically, we encapsulate the finding and contribution of this study as follows,
\begin{itemize}
    \item We refine model training objectives by introducing a sample importance mining approach, which enables better data utilization in data resource-limited scenario. We also integrate hard negative mining into the re-ranking model training through a learn-from-teacher strategy, which is demonstrated critical to ensure the model's performance. 
    \item Through comprehensive offline analyses, we illustrate the proposed approaches enhance offline recall@100 by 35.2\% and NDCG@3 in the hard negative scenario by 94.4\% compared with the legacy QR modeling. Furthermore, results from human audits indicate that these approaches markedly strengthen the model's ability to discern between relevant and challenging irrelevant query pairs, resulting in a noteworthy improvement in classification AUC from 0.51 to 0.79.
    \item We executed online A/B tests in two distinct use cases of QR, including search ranking and search Ads matching, within Amazon product search. The experiment led to a noteworthy overall search revenue gain of +0.14\% and +0.29\% in Japanese and Hindi cases, respectively, and a +0.08\% incremental revenue gain in English case compared to the legacy implementation. The experiment also led to a +0.36\% increase in Ads revenue in the Japanese case. We also observed that the legacy QR solution that effectively works in major markets under English context showed no revenue gains in smaller markets with non-English language. This comparison confirmed the existence of scalability and extensibility issues, and also conclusively demonstrated the effectiveness of the proposed technique in ensuring robustness of QR.
\end{itemize}

\section{Related work}
Web service developers have dedicated substantial effort to query reformulation in order to address the inefficiencies of information retrieval systems \cite{x0}. 

\textbf{Query-query relationship mining} The basis of query reformulation lies in comprehending the connections between distinct queries. Previous research has explored various approaches, such as utilizing Bayesian perspectives \cite{x1,x2,x3} and graph-based approaches \cite{boldi2008query, boldi2009query, beeferman2000agglomerative} to mine query-query correlations leveraging query-document click-through data, which lay the groundwork for query rewriting using customer interaction data. This type of methods has been further enhanced with entropy information to counteract the bias towards popular queries \cite{deng2009entropy}. In this work, we also use query-document data for query similarity mining, but adopt new metrics based on distributional divergence customized for different training stages in QR. Another thread of work \cite{x4, cao2021automated} utilized search log data to establish query-to-query affinity through session-based concepts. Yet, this approach suffers from rule-based restrictions and has a limited usage. 

\textbf{Query reformulation pipeline} With the advent of transformer-based text encoders, there has been a paradigm shift toward leveraging the semantic information encoded within queries for rewrite tasks \cite{x9,x10}, employing concepts like semantic textual similarity and dense text retrieval. This entails generating representative query embeddings, enabling the establishment of query-to-query relationships on a large scale in a computationally efficient manner. While there has been a notable effort to develop QR modeling within E-commerce applications \cite{li2022query,qiu2021query}, these endeavors did not specifically focus on enhancing the behavioral signal of less common queries, and thus, the design of these models was not tailored accordingly. A previous work \cite{2} bridges this gap for Amazon product search with a query reformulation pipeline. In this work, we focus on exploring the robustness issue in the application of QR methodologies, and more specifically, refining the design and implementation of this pipeline to benefit smaller or emerging businesses in regions characterized by multilingual or complex language contexts.

\textbf{Query reformulation with LLM} A more recent idea of QR is to acquire related queries using broad, open-domain knowledge rather than relying solely on in-domain customer data \cite{x12, x13, x14}, which has become feasible due to the rise of large language models (LLMs). The implementation falls into two categories: A) use LLMs in the inference stage following a traditional generative paradigm for QR \cite{x5,x6,x7,x8}, e.g., query expansion and query rewrite. This approach has been adopted in E-commerce search to improve recall of long-tail queries \cite{new1}. B) use LLMs for training data augmentation \cite{new2}, e.g., generate hard negatives. However, the first approach does not fit the purpose of behavioral signal enrichment as it is hard to force rewritten query to have rich behavioral signal. And we will explore how LLM-based data augmentation could dynamically integrate with the hard negative mining used in this paper in future work.

\section{Problem formulation}

In this section, we present a formulation of the query reformulation problem within the behavior enrichment task. We begin with a set of search queries, denoted as $Q$, collected over a specific time frame. We divide this set into distinct subsets: $Q = Q_r + Q_i$, where $Q_r$ denotes behavior-rich queries, $Q_i$ denotes behavior-impoverished queries.

For the Query Reformulation (QR) problem, our objective is to establish mappings for each behavior-impoverished query, $q_{i}\in Q_{i}$, such that $q_{i}:\mapsto \{q_{i1},...,q_{iN}\}$, where $q_{ik}\in Q_{r}$, and each query mapping $(q_{i},q_{ik})$ is considered to be semantically relevant.

\section{Method}
To ensure both computational efficiency and accuracy of QR inference, we follow the paradigm introduced by previous research \cite{2} and decompose the computation into two stages. In the initial retrieval stage, a bi-encoder \cite{7} model is employed to generate embeddings for each query in both $Q_r$ and $Q_i$, constructing a small candidate pool through K-nearest neighbor (KNN) search. In the following re-ranking stage, a cross-encoder \cite{8} model estimates relevance score $s$ for all pairs of a query and its corresponding candidates, based on which only relevant candidates are retained.

In this section, we elaborate on the enhancements introduced to the training of models mentioned above. Our refinements encompass the model training objective, the data mining process to generate signals for training, and an integration of Approximate Nearest Neighbor Negative Contrastive Estimation (ANCE) \cite{5} into the re-ranking model training processes. 

\begin{figure}[h]
  \centering
  
  \includegraphics[width=\linewidth]{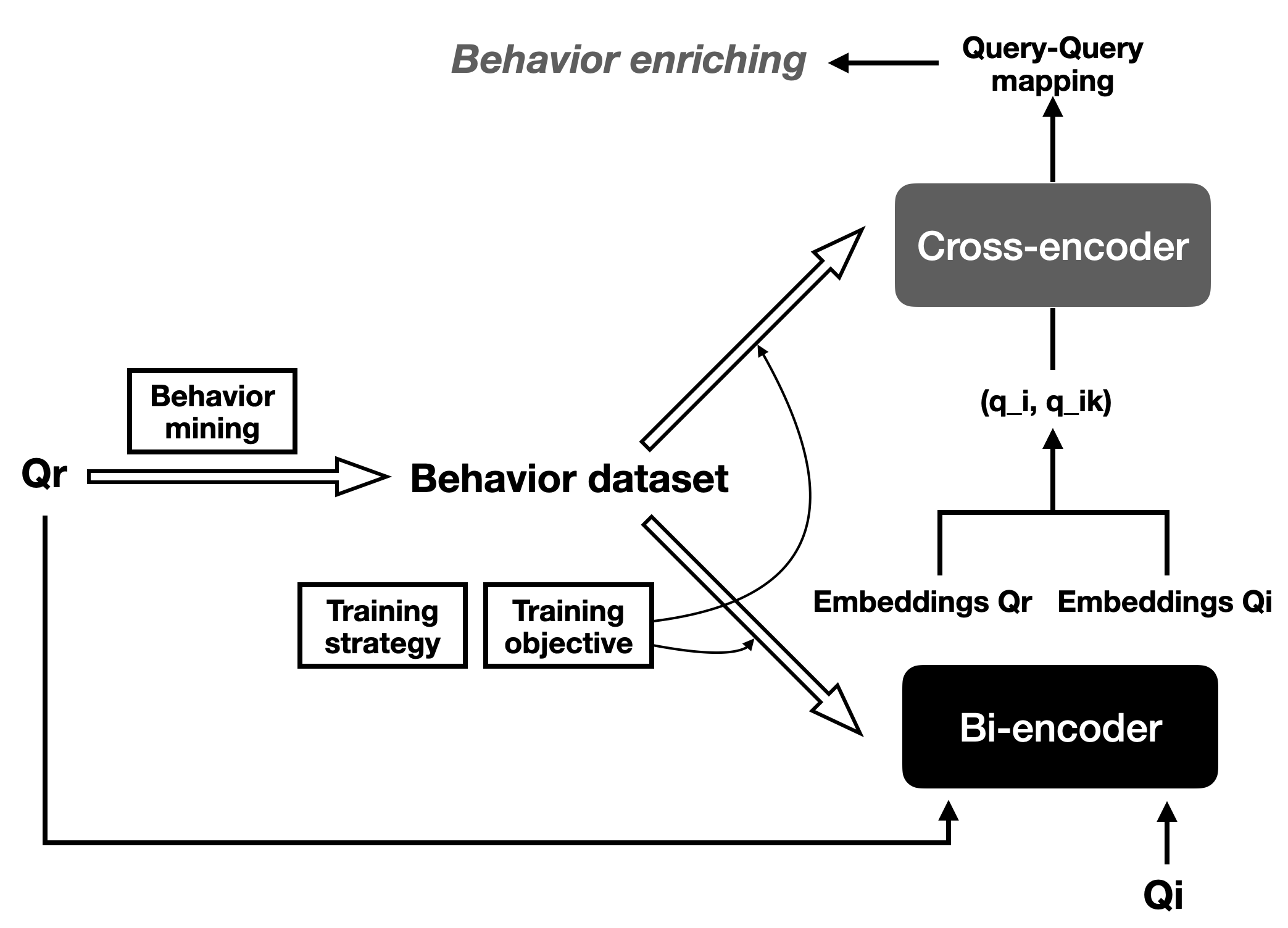}
  \caption{An illustration of a) QR pipeline in model training stage (hollow arrows), b) QR pipeline in online inference system (solid arrows), c) data artifact used in training and inference (text box without borderline), d) our refinements to the QR pipeline (text box with borderline).}
  \label{fig:contribution}
\end{figure}

An illustrative representation of the QR pipeline in both model training stage and in online inference system is shown in Figure \ref{fig:contribution}. We highlight the refinements to model training proposed in this paper with bordered text boxes.

\subsection{Retrieval model training schema}

We use $s_{BE}(q_{i}, q_{j})$ to represent the cosine similarity between L2 normalized embedding of $q_i$ and $q_{j}$. For retrieval task, we define the training objective under the contrastive learning with in-batch negative framework as

\begin{equation}
\mathcal{L}_{\mathrm{retrieval}}=-1/N\cdot\Sigma_{i,k}I(q_i,q_{ik})\cdot \frac{\exp{s_{BE}(q_{i}, q_{ik})/\tau}}{\Sigma_{j\in \mathrm{batch}}\exp{s_{BE}(q_{i}, q_{j})/\tau}}
\end{equation}

\noindent where $N$ is the number of positive query pairs, $\tau$ is a temperature parameter. Specifically, we add a sample importance term $I(q_i, q_{ik})$ to the infoNCE \cite{6} (NT-XENT \cite{5}) objective. We first explain the necessity of adding this term and leave the detailed calculation to the next subection.

\noindent \textbf{RQ1: Why is sample importance critical for the QR solution in the resource-limited scenario?}

\noindent \textbf{Fact1.1:} In the context of QR, the positives are derived from behavioral data, which are prone to being false positives. For instance, the queries “furniture” and “dining table” share numerous associated products, and are usually considered related in many behavioral mining algorithms. However, borrowing behavior of the query “furniture” to the query “dining table” would inevitably relate products like “TV cabinet” to “dining table”, and consequently, significantly compromises search relevance.

\noindent \textbf{Fact1.2:} In resource-limited scenario, simply remove semi-relevant query pairs could harm the training efficacy. As we will show later in the experiment section, in a setting where we only kept top 30\% percent of positive query pairs to ensure a high-quality training dataset, the experiment succeeded in the English case where the training data is abundant, but failed for languages with a relatively insufficiently large dataset. By introducing sample importance, on one hand, in the presence of abundant training data, the impact of less pertinent pairs is mitigated, allowing for more focused model learning. On the other hand, in scenarios with limited supervised resources, pairs that may not be strictly relevant-yet still bear some relevance-can still make a valuable contribution to the model's training.

\subsection{Sample importance through behavioral mining}

This subsection introduce a data mining strategy to derive sample importance for each positive query-query pair. The initial step involves extracting the behavior data (i.e., products and their corresponding purchase counts) of each query $q_i$, denoted as:

\begin{equation}
  B(q_{i}): \{p_{i1}:n_{i1},\ p_{i2}:n_{i2},...\}
\end{equation}

\noindent where $p_{ik}$ represent a product which has been purchased under query $q_i$ and $n$ signifies the corresponding number of purchase. 

We define the query-query relevance measurement as follows.
\begin{enumerate}
    \item Normalize the number of purchase in $B(q_i)$ as $\hat{n_{ik}}=n_{ik}/\sum_{m=1}^{x}n_{im}$.
    \item Calculate Jensen–Shannon divergence (JSD) between normalized purchased product count distributions $\{\hat{n_{11}}, \hat{n_{12}}, ..., \hat{n_{1x}}\}$ and $\{\hat{n_{21}}, \hat{n_{22}}, ..., \hat{n_{2y}}\}$. 
\end{enumerate}

This definition is an improvement based on a prior study\cite{2}, where the query relevance is defined as 
\begin{equation}
  \dfrac{|PP(q_1)\cap PP(q_2)|}{\mathrm{Min}(|PP(q_1)|, |PP(q_2)|)} \cdot \dfrac{|PP(q_1)\cap PP(q_2)|}{|PP(q_1)\cup PP(q_2)|}
\label{eq:score}
\end{equation}

\noindent with $PP(q)$ representing the set of products purchased under query $q$. 

\begin{figure}
     \centering
     \begin{subfigure}[b]{0.23\textwidth}
         \centering
         \includegraphics[width=\textwidth]{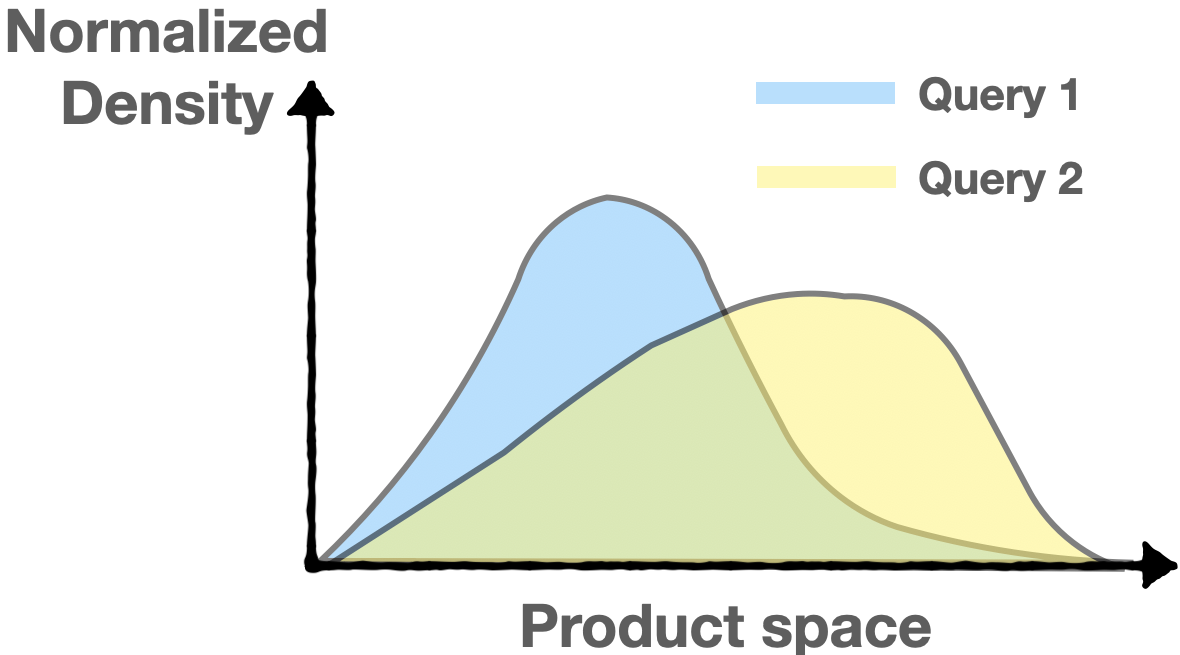}
         \caption{Queries with different semantic meaning}
         \label{fig:asin1}
     \end{subfigure}
     \hfill
     \begin{subfigure}[b]{0.23\textwidth}
         \centering
         \includegraphics[width=\textwidth]{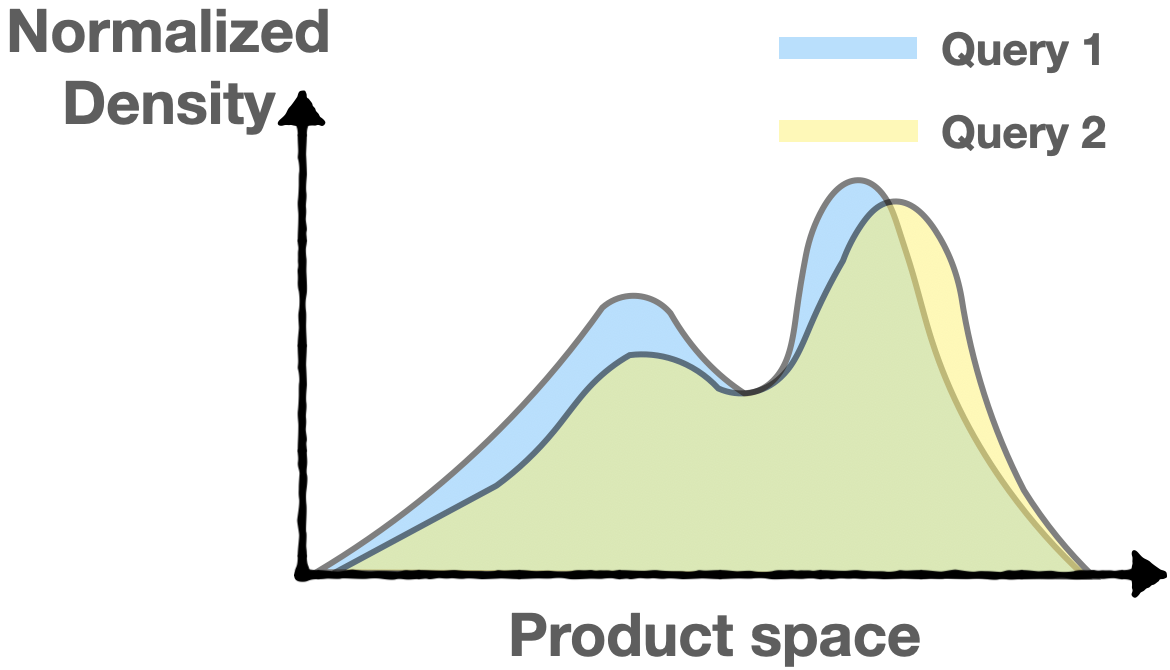}
         \caption{Queries sharing identical semantic meaning}
         \label{fig:asin2}
     \end{subfigure}
    \caption{The distribution of number of purchase on product under different queries.}
\end{figure}

The proposed measurement alleviates the issue that a pair might receive a high relevance score despite having significantly different distributions in their product purchase counts (as shown in Figure \ref{fig:asin1}). Under the proposed measurement, only pairs that share not only the same set of purchased products but also a similar distribution can achieve a high similarity score (as shown in Figure \ref{fig:asin2}). 

To illustrate the impact of this enhancement, we provide examples of the top-1 similar query pairs $(q_i, q_{i1})$ under both measurements method in Table \ref{tab:1}. It becomes evident that the proposed measurement yields more suitable query pairs and eliminates undesirable cases, such as "google pixel watch" and "pixel watch case."

\begin{CJK}{UTF8}{min}
\begin{table}
\caption{Example of top-1 similar query pairs derived with the previous similarity metric (Equation \ref{eq:score}) and the proposed metric based on JS-divergence.}
\begin{tabular}{|p{2.5cm}|p{2.5cm}|p{2.5cm}|}
 \hline
 \rowcolor{arsenic}
 \textcolor{white}{$q_i$} & \textcolor{white}{$q_{i1}$ (previous)} & \textcolor{white}{$q_{i1}$ (proposed)}\\
 \hline
 google pixel watch& pixel watch case & pixel watch \\
 \hline
 usb bluetooth 5.0 & bluetooth tplink & tp-link bluetooth usb bluetooth 5.0 \\
 \hline
 pdp controller & pdp &pdp controller switch\\

 \hline
\end{tabular}
\label{tab:1}
\end{table}
\end{CJK}

\subsection{Re-ranking model training schema}
\label{sec:jsd}
\noindent \textbf{RQ2: What benefit does the re-ranking stage bring under the QR context?}

\noindent \textbf{Fact2:} We take into account the asymmetrical nature of $(q_i, q_{ik})$ pairs in the training data. It is noteworthy that $(q_i, q_{ik})$ forms a suitable pair in the context of QR does not necessarily imply that $(q_{ik}, q_i)$ is also an appropriate pair. In the KNN search-based retrieval task, it is infeasible to finetune the model to differentiate between the positions of queries within a pair, as the embedding calculation only takes one query at a time as input. However, in the re-ranking task, the cross-encoder model can be trained in a position-aware mode, which enables the re-ranking stage to further mitigate false-positives in the end-to-end data generation process. 

Specifically, we use separate similarity measures for retrieval and re-ranking tasks to fully exploit the inductive bias of each model and ultimately address relevance defects. Consider the calculation\begin{multline*}
JSD(dist(q_i)||dist(q_{ik})) \propto \\
KLD(dist(q_{ik})||dist(m_{(i,ik)})) + KLD(dist(q_i)||dist(m_{(i,ik)})))
\end{multline*}
\noindent where $dist(m_{(i,ik)}) = (dist(q_{ik})+dist(q_i))/2$. The first term emphasizes the distribution divergence with respect to the space defined by products associated to $q_{ik}$, which is more sensitive on false-positive cases where major products that compose $dist(q_{ik})$ is out of the distribution represented by $dist(q_i)$; while the second term focuses on the other way around, potentially leading a departure from identifying false positives. Therefore, we only retain the first term of JSD in similarity measurement for the re-ranking task so that the result is contingent solely on whether the products associated with $q_{ik}$ are related to $q_i$. 

\subsection{Query normalization for denoising}

\begin{CJK}{UTF8}{min}
In developing our QR framework for non-English and low-traffic markets, we employ a text-normalization pipeline of preprocessing steps, including stop word removal, stemming, script type normalization (e.g., translation between different spellings of the same word, as illustrated by "こども" to "子供" in the Japanese case), tokenization (e.g., converting "子供マスク不織布" to "子供\_マスク\_不織布"), and token sorting (e.g., reordering "子供\_マスク\_不織布" to "マスク\_子供\_不織布"). It is imperative to note that this process excludes the handling of identified named entities, such as brands or media titles, in order to prevent any inadvertent introduction of defects. 
\end{CJK}

\noindent \textbf{RQ3: Why is query normalization necessary?}

\noindent \textbf{Fact3.1 The data quality perspective:} It has come to our attention that the similarity scores (i.e., the sample importance used for model training) derived from behavioral data is noisy. Within a group of queries with the same shopping intent but slightly different representations, the similarity score between each pair of queries could have a discrepancy. We illustrate this phenomenon in Table \ref{tab:2}, where we present a group of queries all centered around the same shopping intent. We show the similarity scores computed between the aggregated behavior of group and the individual behavior of each query within the group. The queries in this group only differ in terms of tokenization (i.e., the presence or absence of space) and token order, but the similarity scores range from 0.18 to 0.71. Hence, we contend that clustering query behavior through normalization is a necessary step to acquire similarity scores that more accurately reflect the semantic distinctions between queries. 

\noindent \textbf{Fact3.2 The data quantity perspective:} In the construction of positive query-query pair training dataset, we use a filtering step aimed at reducing noise. This step exclude query-product pairs with a low number of purchases. As a consequence, a considerable number of relevant query-query pairs are eliminated due to the sparsity of behavioral signals. The introduction of query normalization offers a significant mitigation to this issue. Specifically, by grouping queries that are identical after normalization and combining their behavioral signals, we can enable a greater number of query-product pairs to surpass the filtering threshold. In this study, it substantially increases the count of query-query pairs from about 9 million to about 348 million. Furthermore, it also expands the number of unique queries with low request traffic (i.e., contributing to the last one-third of traffic) covered by the behavioral dataset, elevating it from about 166 thousand to more than 3 million. (Please note that this number is derived from the entire query set rather than $Q_r$.)

\begin{table}
\renewcommand{\arraystretch}{1.2}
\centering
\begin{CJK}{UTF8}{min}
\caption{An example in Japanese of query similarity disparity within a group. The queries have the same normalized representation but have varied relevance score.}
\label{tab:2}
\begin{tabular}{ |c |c| c| }
    
 \hline
 	
 \rowcolor{arsenic}
 \textcolor{white}{Query} & \textcolor{white}{Translation} & \textcolor{white}{Relevance Score} \\
 \hline
 \colorbox{gray!30}{子供}\_\colorbox{gray!10}{マスク\_不織布} & \colorbox{gray!10}{non-woven} & 0.18\\
 \cline{1-1} \cline{3-3}
 \colorbox{gray!10}{マスク\_不織布}\_\colorbox{gray!30}{子供} & \colorbox{gray!10}{fabric} & 0.71\\
 \cline{1-1} \cline{3-3}
 \colorbox{gray!10}{マスク}\_\colorbox{gray!30}{子ども}\_\colorbox{gray!10}{不織布} &  \colorbox{gray!10}{masks for}& 0.44\\
 \cline{1-1} \cline{3-3}
 \colorbox{gray!10}{マスク\_不織布}\_\colorbox{gray!30}{こども}  & \colorbox{gray!30}{children} & 0.30\\
 \hline
 
\end{tabular}
\end{CJK}
\end{table}

\subsection{Model training with hard negative mining}

Previous research has introduced the concept of Approximate Nearest Neighbor Negative Contrastive Estimation (ANCE) \cite{5} to address the inefficacy of using in-batch negative in contrastive learning. ANCE entails the incorporation of hard negatives, which are selected from the entire dataset for each query within a batch. This selection process is carried out through a sequence of retrieval model inferences and KNN indexing. In the context of QR, we define the set of hard negatives as the queries that are retrieved for $q_i$ but do not share any co-purchases with $q_i$.

\begin{figure*}[h]
  \centering
  \includegraphics[width=0.9\linewidth]{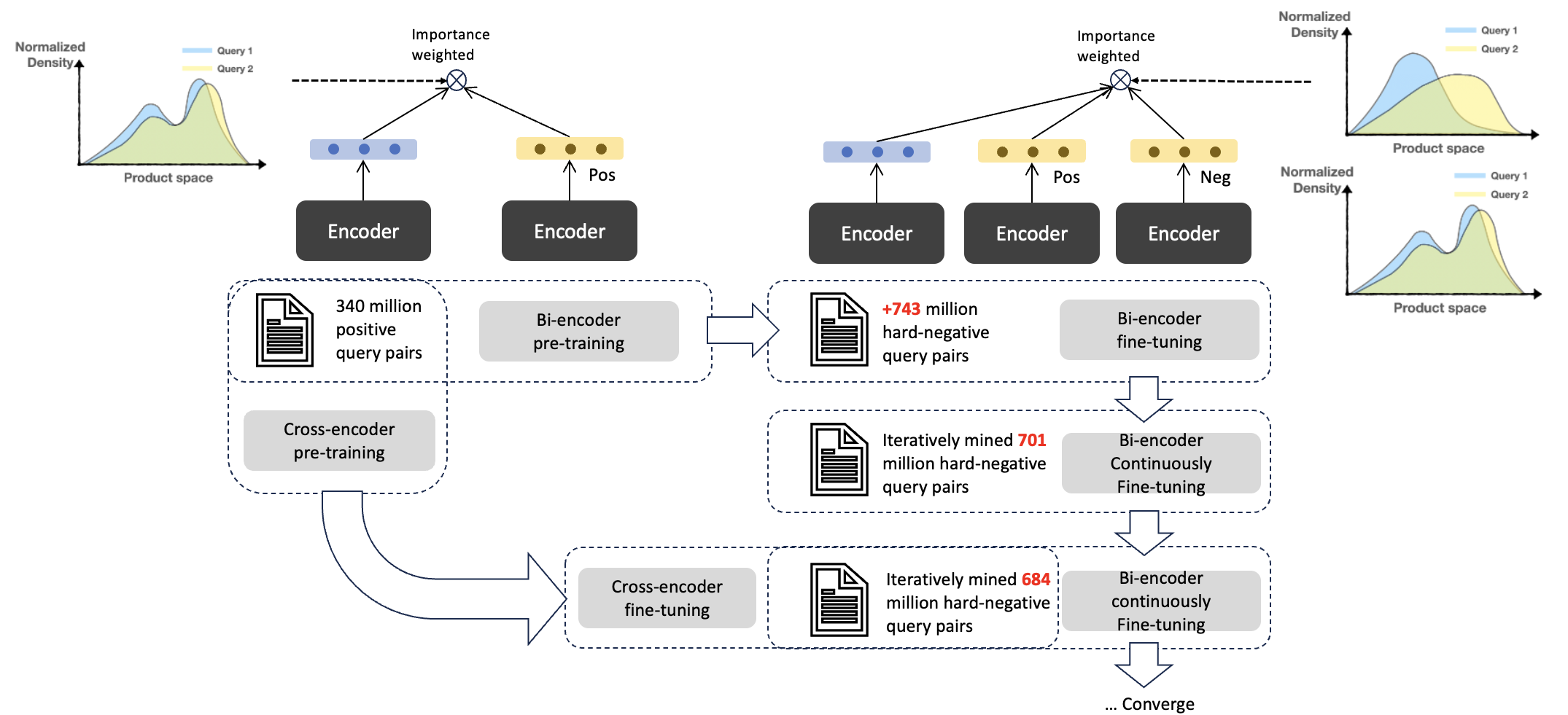}
  \caption{The ANCE mining and model continuous finetune process. Specifically, the bi-encoder model is iteratively finetuned using ANCE data derived by self-retrieved pairs subtracting pairs with co-purchase. Then, the cross-encoder model is finetuned with the ANCE data from the bi-encoder.}
  \label{fig:pipeline}
\end{figure*}

We put forth distinct strategies to harness ANCE in the training of both retrieval and re-ranking models. For the retrieval model, we employ a self-learning strategy, wherein the bi-encoder is iteratively finetuned with the ANCE dataset it generates. Conversely, for the re-ranking model, as it is hard to acquire hard-negatives by cross-encoder itself through a retrieval process, we adopt a learning-from-teacher strategy. Under this approach, the cross-encoder is finetuned using the ANCE dataset generated by the bi-encoder model. We will highlight in the experiment section that this approach is necessary regarding re-ranking model's performance. In addition, we group the pairs associated to each $q_i$ to form a positive set and employ pairwise loss rather than pointwise loss. The necessity for training in a pairwise manner arises particularly in the case of hard negatives, where positive and negative samples often exhibit only nuanced differences in their text representations. Pairwise training can enable the model to capture the supervised signal with greater precision, reduce variance during training, and thus, enhance the overall training efficacy. Specifically, we use circle loss \cite{9} as training objective.
\begin{multline}
\mathcal{L}_{\mathrm{re-ranking, hard}}=\log(1+\Sigma_{j\in H(q_i)}\exp s_{CE}(q_{i}, q_{j}))\cdot \\
(1+\Sigma_{q_{ik}}\exp (-s_{CE}(q_{i}, q_{ik})))
\end{multline}
\noindent where $s_{CE}(q_{i}, q_{j})$ represents the similarity score given by the cross-encoder model, and $H(q_{i})$ represents the hard negatives of $q_i$. The ANCE mining and model continuously fine-tuning process is described in Figure \ref{fig:pipeline}.

\section{Experiment}

In this section, we present our experimental setup and results. Our experiments were conducted in the Amazon product search, with the primary goal of assessing the scalability and extensibility of our query reformulation techniques. In addition to offline evaluations, we tested our proposed QR solution in two downstream online applications to improve behavior-based search ranking features and optimize search Ads matching through A/B testing.
\subsection{Dataset generation}
We construct training and testing dataset with one-year anonymized Amazon search data. We first extract query-product pairs from anonymized historical customer behavior data. Then we perform self-join of these query-product pairs on product, and aggregate the joined result by product to obtain the query co-purchase data. At last, we select 5,049 unique queries from $D$ and extract all query pairs containing the selected queries as testing data (abour 371K). The remaining pairs are divided into training and validation data, using a 90\%/10\% ratio.


\subsection{Offline evaluation criteria}

We employ distinct evaluation metrics for the retrieval and re-ranking tasks. For retrieval, we calculate the recall@100 for both the top-3 paired queries and all queries paired to the given query in the testing data. This separation into top-3 and all queries is motivated by their significance in reflecting the model's performance in downstream ranking and matching applications, respectively. Besides, we calculate both micro-recall, which represents the average recall of each query in the testing dataset, and macro-recall, which is derived by dividing the total number of retrieved paired queries by the total number of all paired queries.

For the re-ranking aspect, we evaluate each query using NDCG@3 (Normalized Discounted Cumulative Gain) with the behavior-based similarity score of all paired queries in the testing data. In this evaluation, we also report NDCG@3 for the hard negative scenario, which involves the union of all retrieved queries and paired queries in the calculation. In the hard negative scenario, the score for unpaired queries is set to zero. This evaluation offers a more precise measure of the re-ranking model's performance in downstream applications, as it closely simulates the context of re-ranking followed by retrieval.

\subsection{Offline experiment result}

We train a series of bi-encoder model and cross-encoder model, separately for retrieval and re-ranking tasks. The proposed approaches are progressively added upon a baseline model to study the individual benefit brought by each approach. The datasets used for training were constructed through query normalization-based augmentation. 

\noindent \textbf{Retrieval} - The experimental results are detailed in Table \ref{tab:retrieval}. The baseline model ( \textbf{rt1}) did not employ sample weighting; rather, it used the top 30\% of pairs for each query as training data. For simplicity, we will analyze the findings specifically for recall@100 in relation to the top-3 queries (in bold font), as the results for all metrics follow a similar trend. Notably, the inclusion of the sample importance term elevated recall@100 by 5.4\% ( \textbf{rt2} vs. \textbf{rt1}). This improvement was further augmented by the addition of ANCE, resulting in an incremental gain of 7.9\% ( \textbf{rt3} vs. \textbf{rt2}). Additionally, through successive iterations of the ANCE dataset and continuous model fine-tuning, recall@100 increased to 0.777 in round 2 ( \textbf{rt4}) and 0.798 in round 3 ( \textbf{rt5}). Note that we only perform three rounds of ANCE as we observed marginal benefit with more rounds. The outcomes suggest that the combined utilization of the sample importance in the loss function and ANCE substantially enhances the retrieval model's performance.

\begin{table}
\centering
\caption{Offline evaluation of retrieval models.}
\label{tab:retrieval}
\begin{tabular}{ |c|c|c|c|c|c|  }
\hline
\multirow{3}{*}{} & \multirow{3}{*}{Model}  & \multicolumn{2}{c|}{Recall (top-3)} & \multicolumn{2}{c|}{Recall} \\
    \cline{3-6}
& \multicolumn{1}{c|}{} & \multicolumn{2}{c|}{@100}  & \multicolumn{2}{c|}{@100} \\
\cline{3-6}
& \multicolumn{1}{c|}{} &micro&macro&micro&macro \\

  \hline
 rt1 & Baseline & \textbf{0.5903}&0.5932&0.4701&0.3451\\
 rt2 & +sample importance & \textbf{0.6441}&0.9322&0.5118&0.3788\\
 rt3 & +ANCE (round 1)&\textbf{0.7232}&0.7256&0.6075&0.4638\\
 rt4 & +ANCE (round 2)& \textbf{0.7766}&0.7771&0.6570&0.4876\\
 rt5 & +ANCE (round 3) & \textbf{0.7984}&0.7995&0.6808&0.5159\\

 \hline

\end{tabular}
\end{table}


\noindent \textbf{Re-ranking} - the model performance is presented in Table \ref{tab:reranking}. The baseline model trained with the KLD objective outperformed the best retrieval model (\textbf{rt5}), indicating the necessity of the re-ranking stage under the QR context. Surprisingly, our findings further revealed an unexpected outcome: as the model's focus shifted towards hard negative samples through the incorporation of ANCE and the adoption of circle loss, the model's performance exhibited a gradual decline in the general scenario, while showing an improvement in the hard negative scenario. This led us to hypothesize that an emphasis on hard negatives during training might result in overfitting. Moreover, this finding prompted a reconsideration of how re-ranking models should be evaluated. Relying solely on paired queries for assessment may not suffice, as it risks producing misleading comparisons among different models. This is particularly pertinent because the practical use of the re-ranking model in the end-to-end inference process mirror hard negative scenarios more closely. Our stance is further substantiated by the results of human auditing, detailed in Section \ref{humaudit}. Here, we observed that evaluations in hard negative contexts correlated more closely with the outcomes of relevance auditing, underscoring the practical applicability of our model in realistic settings.

\begin{table}
\centering
\caption{Offline evaluation of re-ranking models.}
\label{tab:reranking}
\begin{tabular}{ |c|c|c|c|  }
 \hline

 \rowcolor{arsenic}
 \textcolor{white}{} & \textcolor{white}{Model} & 
 \textcolor{white}{NDCG@3} & \textcolor{white}{NDCG@3} \\
 \cline{0-3}
  \rowcolor{arsenic}
  &  & \textcolor{white}{(general)} & \textcolor{white}{(with hard negative)} \\

  \hline
 rt5 & Retrieval & 0.81487 & 0.30102 \\
 rr1 & Baseline & 0.89459 & 0.34844 \\
 rr2 & +ANCE & 0.84462 & 0.64612 \\
 rr3 & +pairwise loss & 0.79172 & 0.67733 \\
 \hline

\end{tabular}
\end{table}

\subsection{Offline query-query relevance auditing}
\label{humaudit}
Conducting A/B tests online demands considerably more resources than offline analysis and entails the risk of deploying models of inadequate quality into production. Hence, establishing an interpretable pattern that bridges the gap between offline findings and actual production results is immensely beneficial.

We extracted a sample of 400 query pairs from the retrieval dataset and conducted a human audit. The assessment of models used for experiment on the audit data is detailed in Table \ref{tab:human}. The model performance was measured through the area under the Receiver Operating Characteristic curve (AUROC) and Spearman correlation.

\begin{table}
\centering
\caption{Model evaluation based on human audit.}
\label{tab:human}
\begin{tabular}{ |c|c|c|c|}
 \hline
 \rowcolor{arsenic}
 \textcolor{white}{Model} & \textcolor{white}{AUROC} & \textcolor{white}{AUROC} & \textcolor{white}{Spearman}\\
 \cline{0-3}
 \rowcolor{arsenic}
  \textcolor{white}{} & \textcolor{white}{(strictly relevant)} & \textcolor{white}{(not relevant)} & \textcolor{white}{correlation}\\
  \hline
  \multicolumn{4}{|c|}{Retrieval} \\
  \hline
rt1	& $0.51263$ & $0.53949$ & $0.0373 $ \\
rt5 & $0.78956$ & $0.7481$ & $0.4754 $\\
\hline
  \multicolumn{4}{|c|}{Re-ranking} \\
  \hline
rr1	& $0.71178$ & $0.57035$ & $0.0572$ \\
rr3 & $0.77072$ & $0.71539$ & $0.4254$ \\
\arrayrulecolor{black}\hline
\end{tabular}
\end{table}

From the findings presented in Table \ref{tab:human}, a noticeable distinction between the \textbf{rt1} and \textbf{rt5} models, as well as between \textbf{rr1} and \textbf{rr3} models emerges. The \textbf{rt1/rr1} model showcased an inability to discern between relevant and irrelevant queries in the hard negative scenario, as indicated by both AUROC and Spearman correlation measures, whereas the \textbf{rt5/rr3} model exhibited an opposing performance. This outcome suggests that the proposed techniques fostered robustness of QR through bolstering the model's capacity in accurately modeling query relevance.

In addition, we noticed that the retrieval model rt6 outperformed the re-ranking model \textbf{rr3} in terms of both AUROC and Spearman correlation metrics. We attribute this disparity to the distinct training objectives of the two models (i.e., JSD for retrieval and KLD for re-ranking). Furthermore, our auditing protocol is more in tune with the requirements of retrieval, which might explain the observed differences in performance. Please refer to Section \ref{sec:jsd} for a detailed exposition.

\subsection{Online experiment}

In this section, we report the A/B test result of QR applications in Amazon product search. We focus on two applications. One is to enhance the product search ranking feature. Specifically, we aim at the customer behavior-based query-product level feature, denoted by $f(p,q)$. With a mapping $q_{i}:\mapsto \{q_{i1},...,q_{iN}\}$ given by QR and $f(p,q)$, we augment the targeted feature by 
\begin{equation}
    \hat{f(q_i,p)}=\alpha f(q_i,p)+(1-\alpha)\cdot \beta \cdot\frac{1}{N}\Sigma_{k}f(q_{ik},p),
\end{equation}

\noindent where $\alpha$ and $\beta$ are hyper-parameters. In this application, we conducted two-week A/B tests in different markets, characterized by Japanese, Hindi, and English, respectively, to evaluate the QR implementation introduced in this study. The results are detailed in Table \ref{tab:ab}, revealing that the proposed implementation resulted in a considerable improvement in both revenue and conversion in all tested markets. By comparing the previous implementation versus the new implementation in the Japanese, Hindi (i.e., \textbf{rt5+rr3} row verses \textbf{rt1+rr1} row), and English (i.e., \textbf{rt5+rr3} row) cases, we can further assert that the proposed refinements to model training are pivotal components for the success of QR practice in terms of scalability and extensibility.

The second application pertains to product search Ads matching. In our experiment performed in Japanese market, we included the Ads matched for $\{q_{i1},...,q_{iN}\}$ directly into the match set of $q_i$. The experiment encompassed both QR implementations, one with the proposed techniques (\textbf{rt5+rr3}), and another without them (\textbf{rt1+rr1}) through separate two-week A/B tests. Table \ref{tab:ab} presents the results, indicating that the model with hard negative-tailored adjustments exhibited a noteworthy gain in both Ads revenue and click-through-rate (CTR), while the model without these adjustments showed only a marginal gain. This outcome strongly suggests that effectively handling hard negatives is a crucial component for the successful practice of QR.

\begin{table}
\renewcommand{\arraystretch}{1.1}
\centering
\caption{Online A/B test results of different applications on markets characterized by different languages.}
\label{tab:ab}
\begin{tabular}{ |c|c|c|c|c|}
 \hline
 \rowcolor{arsenic}
 \multicolumn{5}{|c|}{\textcolor{white}{Product search ranking}} \\
 \hline
 \rowcolor{gray2}
 Language & Control & Treatment & Revenue & Conversion \\
 \hline
\multirow{2}{*}{Japanese} & w/o QR & rt1+rr1 & $-0.05\%$ & $-0.16\%$ \\
& w/o QR & rt5+rr3	& $+0.14\%$ & $+0.08\%$ \\
 \hline
\multirow{2}{*}{Hindi} & w/o QR & rt1+rr1	& $+0.08\%$ & $+0.00\%$ \\
& w/o QR & rt5+rr3	& $+0.29\%$ & $-0.01\%$ \\
 \hline
English & rt1+rr1 & rt5+rr3	& $+0.08\%$ & $+0.06\%$ \\
 \hline
 \rowcolor{arsenic}
 \multicolumn{5}{|c|}{\textcolor{white}{Product search Ads matching}} \\
 \hline
 \rowcolor{gray2}
Language & Control & Treatment & Ads Revenue & Ads CTR \\
\hline
\multirow{2}{*}{Japanese} & w/o QR & rt1+rr1	& $+0.06\%$ & $+0.01\%$ \\
& w/o QR & rt5+rr3	& $+0.36\%$ & $+0.36\%$ \\
 \hline
\end{tabular}
\end{table}
Additionally, we would like to clarify that the experiments conducted for both applications only introduced changes for queries with low traffic, specifically for those representing the last one-third of overall traffic in Amazon's product search. The reported numbers were derived from the entire traffic analyzed by Amazon's in-house A/B test system. As a result, the actual impact on queries with low traffic is anticipated to be more substantial than what is being reported.

\section{Conclusion}

In this study, our primary focus centers on query reformulation (QR) as a remedy to address the limited behavioral data available for queries with historically low customer engagement within the realm of E-commerce. We have identified inherent issues within the existing QR solution, particularly emphasizing challenges related to its scalability and extensibility when applied to smaller markets that lack adequate training data and possess unique language characteristics.

Our investigation was prompted by the inability to replicate the success of the existing QR implementation, proven effective in large Enligsh-markets, in other smaller or emerging markets. To resolve these issues, we have introduced various approaches aimed at enhancing the training of models used in the QR pipeline. These enhancements were focused on refining behavior data mining, as well as optimizing the training strategies and objectives of the models. Through comprehensive A/B testing on two online applications in Amazon product search - specifically, product search ranking and product search Ads matching - we confirmed the necessity and efficacy of these approaches.

\bibliographystyle{ACM-Reference-Format}
\bibliography{sample-base}


\end{document}